\documentclass[conference]{IEEEtran}
\usepackage{nicematrix}
\usepackage{float}
\usepackage{color}
\usepackage{cite}
\usepackage{textcomp}
\usepackage{xcolor}
\usepackage{balance}
\usepackage{graphicx}
\usepackage[english]{babel}
\usepackage{amsthm,amssymb}
\usepackage{subcaption}
\usepackage{mathtools}
\usepackage{tikz}
\usepackage{pgf}
\usepackage{amsfonts}
\usepackage{bm}
\usepackage{todonotes}
\usepackage[bottom]{footmisc}
\usepackage[ruled,linesnumbered]{algorithm2e}
\usepackage[T1]{fontenc}
\usepackage{bbm}
\usepackage{enumerate}

\usepackage{setspace}
\usepackage{array}
\usepackage{multirow}
\usepackage{diagbox}
\newtheorem{definition}{Definition}[]
\allowdisplaybreaks
\def\BibTeX{{\rm B\kern-.05em{\sc i\kern-.025em b}\kern-.08em
    T\kern-.1667em\lower.7ex\hbox{E}\kern-.125emX}}
\begin{document}
%\doublespacing
\title{Geometry Based UAV Trajectory Planning for Mixed User Traffic in mmWave Communication}

\author{\IEEEauthorblockN{Sk Abid Hasan, Lakshmikanta Sau, and Sasthi C. Ghosh}

\IEEEauthorblockA{Advanced Computing \& Microelectronics Unit\\
Indian Statistical Institute, Kolkata 700108, India\\
Emails: abidsekh.personal@gmail.com, lakshmikanta\_r@isical.ac.in,  sasthi@isical.ac.in}}
\maketitle
\begin{abstract}
Unmanned aerial vehicle (UAV) assisted communication is a revolutionary technology that has been recently presented as a potential candidate for beyond fifth-generation millimeter wave (mmWave) communications. Although mmWaves can offer a notably high data rate, their high penetration and propagation losses mean that line of sight (LoS) is necessary for effective communication. Due to the presence of obstacles and user mobility, UAV trajectory planning plays a crucial role in improving system performance. In this work, we propose a novel computational geometry-based trajectory planning scheme by considering the user mobility, the priority of the delay sensitive ultra-reliable low-latency communications (URLLC) and the high throughput requirements of the enhanced mobile broadband (eMBB) traffic. Specifically, we use geometric tools like Apollonius circle and minimum enclosing ball of balls to find the optimal position of the UAV that supports uninterrupted connections to the URLLC users and maximizes the aggregate throughput of the eMBB users. Finally, the numerical results demonstrate the benefits of the suggested approach over an existing state of the art benchmark scheme in terms of sum throughput obtained by URLLC and eMBB users.
\end{abstract}
\vspace{-.5em}

\section{Introduction}\label{intr}
\vspace{-.5em}

\noindent With the development of communication technologies, wireless communication is becoming the backbone of daily life. Because of their low acquisition costs, ease of deployment, and hovering capabilities, unmanned aerial vehicles (UAVs) are seen as a promising solution to meet ever-increasing traffic demands, improve coverage area, and avoid obstacles \cite{survey}.
UAV-assisted millimeter wave (mmWave) communication is an appealing option to meet these growing demands because of its high bandwidth and an enormous amount of unoccupied spectrum. However, due to very high frequency, mmWaves suffer from its own set of shortcomings like significantly high penetration and propagation losses,
% requiring a line of sight (LoS) for effective communication.
resulting in connection failure in the presence of obstacles\cite{mmw, drams_lakshmi_da}.
According to the data requirements and delay sensitivity, there are mainly three different types of user traffic: enhanced mobile broadband (eMBB), ultra-reliable low-latency communications (URLLC), and massive machine-type communications (mMTC) \cite{traffic, mmtc}. Since mMTCs is about wireless connectivity to tens of billions of machine-type terminals, here, we consider cost-efficient URLLCs and eMBB traffic that is adequate for human-type cellular communications \cite{mmtc}.  The main focus of this manuscript is to develop a UAV trajectory planning that takes into account the user's traffic characteristics and mobility. 
%, but in case of mmWave communications due to higher path-loss, the NLoS links are almost infeasible\cite{drams_lakshmi_da}.

In UAV-assisted air-to-ground (A2G) communication, direct line of sight (LoS) links may be blocked due to the presence of obstacles. The authors in \cite{3d}, proposed a path loss model for both LoS and non-LoS (NLoS) communication. The authors in \cite{LAP}, use a sigmoid function to model the probability of LoS to a user, characterize the UAV's coverage radius as a function of path loss, and show how optimizing the coverage radius in an interference-free environment can maximize the UAV height. The coverage analysis for low-altitude UAV networks in urban areas has been investigated well in \cite{Uav_cov}.
However, both \cite{LAP} and \cite{Uav_cov} considered that the users are static. 
In \cite{placement}, the authors proposed an optimal UAV placement technique with respect to power loss and overall outage by considering both users and obstacles to be static.  However, in most of the practical scenarios, users may not be static in nature \cite{rwp}. Moreover, in \cite{placement1}, the authors proposed an optimization-based UAV placement strategy by considering the achievable data rate of the users. However, they did not consider the presence of obstacles and assumed LoS links are always available to the ground users. Furthermore, the authors of \cite{power} and \cite{energy} gave a beautiful insight into power control and energy-efficient trajectory planning. However, they also did not consider the aspect of user mobility. A joint power control and optimal placement techniques have been proposed in \cite{joint_place} considering static URLLC users only. The authors of \cite{existing} proposed a trajectory planning algorithm without considering the impact of user's traffic characteristics. To the best of our knowledge, this is the first work that considers heterogeneous user traffic characteristics along with user mobility for UAV trajectory planning in the presence of obstacles.

In this manuscript, we propose a UAV trajectory planning technique to serve the users by considering their traffic characteristics and mobility in the presence of obstacles. Specifically, due to the delay sensitivity of the URLLC traffic, we consider that a maximum number of URLLC users must be served. 
Here, we assume that the NLoS links are unable to provide the minimum data rate threshold for URLLC users and hence only LoS links are considered \cite{drams_lakshmi_da}. On the other hand, since eMBB traffic is susceptible to minor disruptions, we maximize the aggregate throughput of eMBB users with the
help of both LoS and NLoS links, while ensuring that the maximum number of URLLC users is being served. Here, we use some basic computational geometry concepts, like the Apollonius circle and minimum enclosing ball of balls, to predict the appropriate UAV location for the next time instant to meet the desired objective. More specifically, our contributions are as follows:

\begin{itemize}
     \vspace{-0.2em}
    \item In order to serve maximum URLLC users, we find the Apollonius circle or the minimum enclosing ball of balls and accordingly the probable zone for UAV location. Thereafter, discretizing that zone, we identify a set of potential candidate locations for UAV positioning, which can provide LoS links to all the probable positions of the URLLC users.

    \item Next, after finding the zone for URLLC users, we move the UAV to a point within that zone from where the eMBB users get maximum aggregate throughput with the help of both LoS and NLoS links. 
    \item If we get more than one such point, we move the UAV to that point for which the displacement with respect to the previous location of the UAV is minimum.
\end{itemize}
We performed extensive simulations to demonstrate the
benefits of the suggested approach over an existing state of the
art benchmark scheme in terms of the sum throughput obtained by URLLC and eMBB users.

This paper is organized as follows. In Section \ref{system}, we discuss the system model. In Section \ref{prop_strat}, we demonstrate the UAV location-finding algorithm. We discuss the simulation results and compare them with an existing benchmark in Section \ref{simulation}. Finally, concluding remarks are given in Section \ref{conclusion}.

\section{System Model}\label{system}
\subsection{Network Topology}\label{net_top}
\begin{figure}
    \centering
    \includegraphics[width=\linewidth]{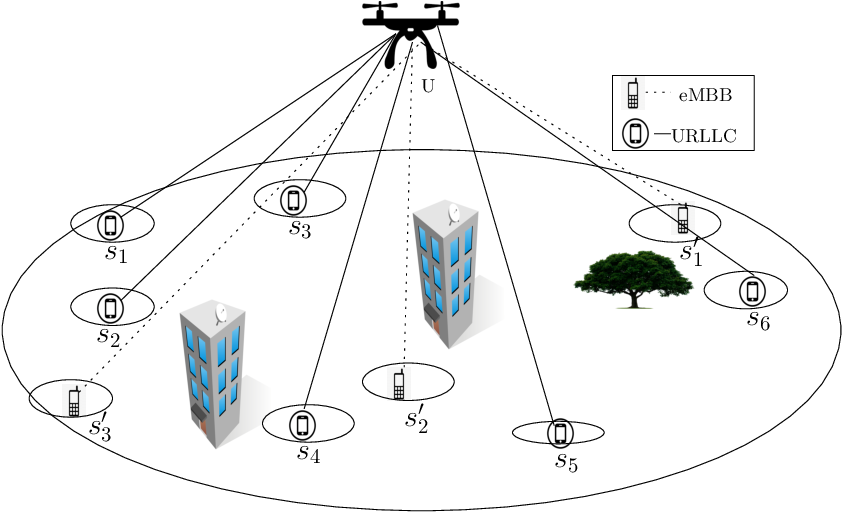}
    \caption{Communication Environment}
    \label{as}
\end{figure}

\noindent We consider an A2G cellular wireless communication system that consists of  $m$ number of user equipments (UEs) $d_1,\cdots, d_m$, $k$ number of UAVs $u_1,u_2,\dots,u_k$, and $n$ number of randomly located static obstacles. 
Additionally, we assume that each UE and UAV consists of a single and $l$ number of antennas, respectively, and all the UAVs are connected with a macro base station (BS) $B$ through a backhaul network. Note that UAVs are designed as multiple aerial BSs that can serve a maximum $l$ number of UEs at a time. However, in a particular time instance, each UE is connected with at most one UAV. Furthermore, we also assume that the coverage area and flying height of a UAV are $R$ and $h$, respectively. The system model is demonstrated in Fig. \ref{as}.

\subsection{Channel Model}
\noindent In our considered topology, UAVs, and UEs are mobile in nature. As a result, due to the presence of static obstacles, the direct LoS may be blocked by the obstacles. Therefore, the transmitted signals from UAV reach UE via the LoS and NLoS links. Due to the high path loss at mmWave frequencies, we consider only LoS links for successful communication. Let us consider that, for a particular $u_i-d_j$ link, depending on the $73$ GHz model, the path loss (in dB) is modeled as \cite{paramss} 
 \vspace{-.5em}
\begin{align}
    PL_{i,j}(d_{i,j})[dB]=\alpha +10\beta \log_{10}(d_{i,j}),
\end{align}
 where $\alpha$ and $\beta$ are environment-specific constants that characterize signal attenuation, and $d_{i,j}$ is Eucledian distance between $u_i$ and $d_j$. Now,
the received power at $d_j$, in $dBm$, is computed using the link budget equation;
\vspace{-.5em}
\begin{align}
    P_{i,j}[dBm]=P_i+G_i+G_j-PL_{i,j}(d_{i,j}),
\end{align}
where $P_i$ is the transmit power, $G_i$ and $G_j$ are the antenna gains of $u_i$ and $d_j$ respectively. 
% So, in linear scale we obtain,
% \vspace{-.5em}
% \begin{align}
%     P_{i,j}^{\rm{lin}} = 10^{P_{i,j}[\text{dBm}]/10}
% \end{align}
 Therefore, the corresponding throughput at user $d_j$ for UAV $u_i$ is \cite{paramss}
 %\vspace{-1.5em}
\begin{align}\label{throughput}
    T_{i,j}=B_w \log_2\left(1+\frac{P_{i,j}\times |g_{i,j}|^2}{\sigma_0}\right),
\end{align}
where $B_w$ is channel bandwidth, $\sigma_0^2$ is the variance of the circularly symmetric zero mean additive white Gaussian noise and $|g_{i,j}|$ is the Rician or Rayleigh random variable depending on whether it corresponds to LoS or NLoS model, respectively\cite{lakshmi_da_priority}. Note that, here, we assume orthogonal frequency division multiple access (OFDMA) where each UE communicates using orthogonal resource blocks \cite{freq_can}.

% Furthermore, the probability of LoS connection between $U_i-d_j$ is
% \begin{equation}
%     P_{\rm {LoS}} = \frac{1}{1 + a \exp(-b( \frac{180}{\pi} \phi_{ij} - a))},
% \end{equation}
% where $\phi_{ij}$ is the elevation angle between $U_i$ and $d_j$, $a$ and $b$ are the constant that depend on the environment.

% Note that the randomly generated channels suffer from large-scale fading and the LoS links follow the Rician distribution \cite{dramp}. Therefore, the aggregate path loss is given by
% \begin{align}
%     h_{\rm arg} = & h_{\rm {LoS}} P_{\rm {LoS}} + h_{\rm {NLoS}} P_{\rm {NLoS}} \nonumber\\
%     &= \frac{A}{1 + a \exp(-b( \frac{180}{\pi} \phi_{ij} - a))} \nonumber \\
%     & \qquad\qquad + 20 \log \left( \frac{r_{ij}}{\cos(\phi_{ij})} \right) + B,
% \end{align}
% where $A = \zeta_{\rm {LoS}} - \zeta_{\rm{NLoS}}$ and 
% $B = 20 \log \left(\frac{4 \pi f_c}{c}\right) + \zeta_{\text{NLoS}}$,considered and $r_{ij}$ is the Euclidean distance between the perpendicular projection of the coordinate of $U_i$ on the $2$-D plane and  $d_j$  \cite{3d}.

\subsection{User Traffic Characterization}\label{user_traf}
\noindent In our considered network topology, according to the user's delay tolerance, we characterize the user traffic in two separate scenarios, namely eMBB and URLLC. Note that URLLC traffic is designed for very low latency and high-reliability applications, enabling real-time mission-critical communication where even a minor delay or packet loss can lead to failures or catastrophic consequences. However, eMBB traffic is designed to provide high-speed, high-capacity, and data-intensive applications. Unlike URLLC, eMBB applications can tolerate minor delays without causing critical failures.
Here we focus on URLLCs and eMBB
traffic that is adequate for human-type cellular and wireless
communication, keeping mMTCs outside the scope of this manuscript.
%Since eMBB \cite{embb} is focused on high-speed data transmission, handoff plays a crucial role in ensuring seamless connectivity and uninterrupted user experience. 

\subsection{UAV and UE Mobility Model}\label{mob_model}
\noindent For a duration of $\Delta t$ time, we assume that both UAVs and UEs can move according to any mobility model with a fixed velocity \cite{rwp}. However, their positions and velocities are at the begining of $\Delta t$. The velocity ranges of $d_i$ and $u_j$ are $v_{i} \in (0,v_{\max})$ and $V_{j}\in (0,V_{\max})$, respectively. Here $v_{\max}$ and $V_{\max}$  are the maximum velocities of UEs and UAVs respectively, where $v_{\max} \leq V_{\max}$. Since the height of UAV is assumed to be remain fixed, during $\Delta t$ time, both $d_i$ and $u_j$ will stay within the disks of radii $v_i\Delta t$ and $V_j\Delta t$, respectively. 
\begin{figure*}
    \centering
    \includegraphics[width=0.67\linewidth]{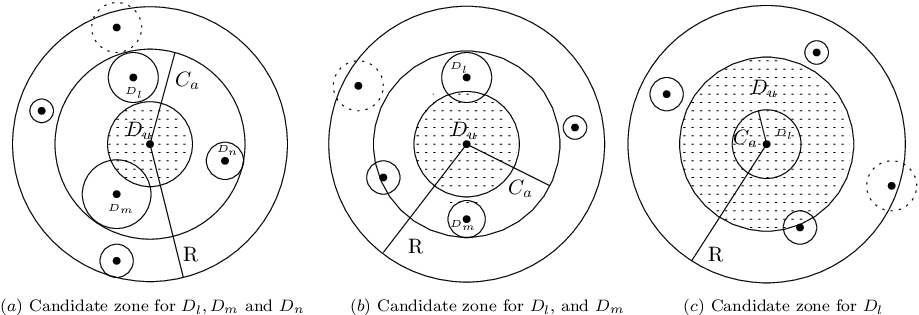}
    \caption{Candidate zone for URLLC traffic}
    \label{apo_dem}
    %\label{fig:enter-label}
    \vspace{-4mm}
\end{figure*}

\section{Proposed strategy}\label{prop_strat}

%According to Section \ref{net_top}, each UAV $u$ can provide services to at most $l$ UEs. Specifically, assume that $S$ and $S'$ are the URLLC and eMBB UEs, respectively, that are being served by a particular UAV $u$ at $t=t_0$, where $|S|+|S'|\leq l$. Due to the movement of the UEs, a few of them may be out of the coverage area of $u$ during $t_0$ to $t_0+\Delta{t}$ time interval. 

\noindent In this section, we propose a trajectory planning algorithm for UAV $u$ based on both URLLC and eMBB traffic. Specifically, first, we propose an algorithm to identify a zone from where the highest number of URLLC users will be covered by $u$. Next, we propose an algorithm to find the final location of $u$ within the identified zone, which serves the maximum amount of eMBB traffic. The complete process is described below.
%\vspace{-.5em}
\subsection{\textbf{Optimal UAV position for URLLC traffic}}\label{URLLCtraf}
\noindent Let $S$ be the set of all URLLC users that are being served by $u$ at a particular time instance $t=t_0$ and their corresponding positions are given by
\begin{equation}
    e_i=(x_i,y_i), \; i=1, \cdots, |S|,
\end{equation}
 where $|S|$ denotes the cardinality of $S$. Additionally, we consider that $e_0=(x_0,y_0)$ is the position of $u$ at $t=t_0$.  Due to the mobile nature of the user $s_i \in S$, according to Section \ref{mob_model}, $s_i$ lies within a disk of radius $v_i\Delta t$ during $t_0$ to $t_0+\Delta t$, where $v_i$ is the velocity of $s_i$. Here, we denote the disk within which $s_i \in S$ can lie, as 
 \begin{align}
 &D_i=\left\{(x,y)\; :\; (x-x_i)^2+(y-y_i)^2\leq v_i\Delta t\right\}.
     % &D_i=\{(e_i, v_i.\Delta t), \;\; i=1,\cdots, |S|.
 \end{align}
 Similarly, $u$ may lie inside a disk $A$ of radius $V\Delta t$, with $e_0$ at its centre, during $t_0$ to $t_0+\Delta t$. That is,
 \begin{equation}
    A=\left\{(x,y)\; :\; (x-x_0)^2+(y-y_0)^2\leq V\Delta t\right\}, 
 \end{equation}
 where $V$ is the velocity of $u$. In our considered topology, both $s_i$ and $u$ move in a $2$-dimensional plane, i.e., $D_i \subset \mathbb{R}^2$ and $A \subset \mathbb{R}^2$. We now formally define when $s_i$ is said to be coverable and covered by $u$.
 \vspace{-.5em}
\begin{definition}{(Coverable):\;\;} An user $s_i\in S$ located at $e_i$ is said to be coverable by $u$ located at $p$, if the Euclidean distance between them, $d(p,e_i) \leq R - (v_i\Delta t)$.
\end{definition}
 \vspace{-.5em}
\begin{definition}{(Covered):\;\;} An user $s_i\in S$ located at $e_i$ is said to be covered by $u$ located at $p$, if 
\begin{enumerate}\label{cover}
    \item $s_i$ is coverable by $u$, and
    \item $\forall \; q \in D_i, \; \exists$ a LoS between $p$ and $q$, and the achievable data rate obtained by $s_i$ located at $q$ from $u$ is greater than the minimum data rate threshold $R_{\rm req}(s_i)$ of $s_i$.
\end{enumerate}
\end{definition}
From the core of computational geometry, if $C$ is a set of $d$-dimensional balls, there exists $E \subset C$ with at most $(d+1)$  balls such that the minimum enclosing ball of $E$ equal to that of $C$, i.e.,  $MB(E) = MB(C)$, where $MB(.)$ denotes the minimum enclosing ball \cite{miniball}. Note that due to the mobility of $s_i$ and $u$, and randomly placed obstacles in the environment, $u$ may not be able to cover all $s_i$ during $t_0$ to $t_0+\Delta t$. In this context, we consider only LoS links for URLLC users for reliability and data sensitivity. More precisely, we assume that the NLoS links are unable to provide the minimum data rate threshold for URLLC users. The {\it Apollonius circle} $C_a$ \cite{apollonius} of three circles is the minimum radius circle that contains all three circles and is tangent to each of them. With the help of $C_a$, we find the region for $u$ that covers a maximum number of $s_i\in S$ as described below.

Let us assume that $D_l$, $D_m$ and $D_n$ be three coverable disks with radii $r_l$, $r_m$ and $r_n$ with respect to $s_l$, $s_m$ and $s_n \in S$, respectively.
Additionally, we also consider that the center and radius of the Apollonius circle $C_a$ correspond to $D_l$, $D_m$ and $D_n$ are $(h,k)$ and $r_a$, respectively, as demonstrated in Fig. \ref{apo_dem}(a).
Note that, according to the construction of $C_a$, the distance between the center of $C_a$ and the center of disk $D_w$ is $r_a-r_w$, where $w=l,m,n$.
Therefore, we get
\vspace{-.5em}
\begin{equation}
    (h - x_w)^2 + (k - y_w)^2 = (r_a - r_w)^2,\;\;w=l,m,n.
\end{equation}
Solving these three equations, we get $(h,k)$ and $r_a$.

%\textcolor{blue}{Now that our primary goal is to find a region for $u$ from where most of the $s_i$ will be covered.} Accordingly, we deliberately find a smaller region and check the covering conditions for each point in it rather than discretizing the entire disc $A$ and verifying the same for each point individually\todo{this part we can state briefly at the beginning of proposed strategy} . Therefore, to find the smaller region, 

We now define a function $A_{\rm ovl}$ that provides an overlapping region between two disks. Let us consider that
\vspace{-.25em}
\begin{align}
 &X_i=\left\{(x,y)\; :\; (x-p_i)^2+(y-q_i)^2\leq r_i\right\}, \;i\in \{1,2\}\nonumber,
 \end{align}
be two disks having centers at $(p_1,q_1)$ and $(p_2,q_2)$ with radii $r_1$ and $r_2$, respectively. Therefore, $A_{\rm ovl}(X_1,X_2)$ can be computed as,
\vspace{-.75em}
\begin{align}\label{eqn_o}
A_{\rm ovl}(X_1,X_2) =& 
r_1^2 \cos^{-1} \left(\frac{d^2 + r_1^2 - r_2^2}{2 d r_1} \right) \nonumber \\
& + r_2^2 \cos^{-1} \left(\frac{d^2 + r_2^2 - r_1^2}{2 d r_2} \right) - \eta
\end{align}
where, $d = \sqrt{(p_2 - p_1)^2 + (q_2 - q_1)^2}\;\;$ and
\begin{align}
& \eta= \prod_{\epsilon_1,\epsilon_2,\epsilon_3 \in \{-1,1\}}\frac{1}{2}\sqrt{\left( \epsilon_1d+\epsilon_2r_1+\epsilon_3r_2\right)}\\
%\vspace{14mm}
& 
\text{ with the condition}\;\epsilon_1+\epsilon_2+\epsilon_3\geq 2.\nonumber
\end{align}
Let $D_u$ be the region with center $(h,k)$ and radius $R-r_a$ that satisfies the following conditions: 
\begin{align}
    &i)\quad R\geq r_a \label{c_condi}\\
    &ii)\quad A_{\rm ovl}(A,D_u)> 0. \label{c_condi1}
\end{align}
Additionally, to determine the position within $D_u \cap A$ from where $D_l, D_m$ and $D_n$ are covered, we discretize $D_u \cap A$ into cells/grids of unit size. 
Let $D_u(l,m,n)$ be the set of all grid centers of $D_u\cap A$ that cover $D_l, D_m$ and $D_n$. That is,
\begin{equation}
    D_u(l,m,n)= \{z \in D_u\cap A:D_l, D_m, D_n\;  \text{covered from}\; z\}.\nonumber
\end{equation}
% Therefore, 
% if $D_i, D_j$ and $D_k$ are covered from all the points of $D_u\cap A$, we will consider that region as a potential region for $U$.}
% Now, let us assume that $D_i, D_j$, and $D_k$ are covered by all the points of  $D_u\cap A$ and rename it as $D_u(i,j,k)$.
Furthermore, by repeating this procedure for every possible triplet $D_l, D_m$ and $D_n$, we can determine the maximum $\binom{|S|}{3}$ number of regions in which we can deploy $u$. 
% Moreover, to determine the position of $U$, we discretize each $D_u(i,j,k)$ into unit cells, where $\; i,j,k=1,\cdots |s|,$ and $i\neq j\neq k $.
Thus the potential candidate zone of $u$ is defined by
\begin{equation}
    Z=\bigcup_{\substack{{1 \leq l,m,n \leq |S|}\\ l\neq m\neq n}}D_u(l,m,n).
\end{equation}

%Consequently, to find the appropriate position $z\in Z$ from where the maximum $s_\in S$ will be covered, we construct a matrix based on their coverage status as follows.
%Let $M$ be a $|Z| \times |S|$ matrix, where $|Z|$ and $|S|$ denote the number of possible cells for $u$ and the number of URLLC users, respectively. Therefore, the matrix $M$ is represented by $M=(M_{ij})$, where\todo{Should not we explain the updation rule for M a bit?}

Consequently, to find the appropriate position $z_j \in Z$ from where the maximum $s_i \in S$ will be covered, we construct a $|S| \times |Z|$ matrix $M=(M_{ij})$ based on the coverage status of $s_i$ with respect to $z_j$, where $M_{ij}$ is as defined below.
\vspace{-.5mm}
\begin{equation}
    M_{ij}=\begin{cases} 
        1, & \text{if } s_i \;\text{is covered from } z_j \\
        0, & \text{Otherwise}.
    \end{cases} 
\end{equation}
Here, $M$ is a binary matrix whose $j$-th column sum is denoted by
\vspace{-2mm}
\begin{equation}
    S(z_j)=\sum_{i=1}^{|S|}M_{ij}.
\end{equation}
Therefore, the column with the maximum number of one provides the location of $u$ from where the maximum number of $s_i$ will be covered. Let us consider that 
\begin{equation}
    S(z)= \max \{S(z_1),\cdots,S(z_{|Z|})\}
\end{equation}
be the maximum column sum corresponding to the cell $z$. Hence, according to our proposed strategy, if we move $u$ at $z$, the maximum number of $s_i$ will be covered. Note that if there exist multiple cells $z$ with highest column sum, we get a set of cells for the same. Consequently, let us consider that $Z_u$ be the set of all the cells that satisfy the following:
\begin{equation}
    Z_u= \{ z_j \in Z \;: S(z_j)=S(z)\}.
\end{equation}

 Note that if for all triplets $D_l$, $D_m$ and $D_n$, $D_u(l,m,n)= \phi$, in a similar way as stated above, we can find $Z_u$ by considering all possible pairs of disks $D_l$ and $D_m$, where $l,m=1, \cdots, |S|$ and $l \neq m$. Moreover, if for all pairs $D_l$ and $D_m$,  $D_u(l,m) = \phi$, we do the same thing for each individual disk $D_l$ and finally get $Z_u$. Note that for the last two cases, we use a minimum enclosing ball $C_{MB}$ with radius $r_{MB}$ instead of using the Apollonius circles, as demonstrated in Figs. \ref{apo_dem}(b) and (c). The complete process is described in Algorithm \ref{URLLC_ALGO}.

 Now in the following subsection, we will find the optimal position  $z \in Z_u$ of $u$, which maximizes the eMBB traffic.

\subsection{\textbf{Optimal UAV position for both URLLC and eMBB:}}
\noindent Unlike URLLC traffic, eMBB traffic can withstand small delays and make the partial coverage scenario considerable. Here partial coverage means that $u$ can serve an eMBB user even if the entire disk concerning the user is not covered from $u$. Here, we allow both LoS and NLoS communication for eMBB traffic. From the previous subsection, we have found $Z_u$, the region from where the maximum number of URLLC traffic is served. Here, we find the region $Z'_u \subset Z_u$ from where the aggregate throughput obtained by eMBB users is maximum as described below.

\begin{algorithm}[h!]
    \caption{Optimal UAV position finding algorithm}
    \label{URLLC_ALGO}  
\KwIn{$S, A, D = \{ D_i \mid s_i \in S \}$}
\KwOut{$Z_u$}
Initialize: $Z=\emptyset$, $Z_u=\emptyset$\;
\For{$D_l,D_m,D_n \in D$ and $l\neq m\neq n$}{
    Compute: $C_a \;\;and\;\; D_u$\;
    \If{$r_a\leq R$ and $A_{ovl}(A,D_u)>0$}{
        Compute: Potential Zone $D_u(l,m,n)$\;
        $Z \gets Z \cup \{D_u(l,m,n)\}$\;
    }    
}
\uIf{$|Z|=0$}{
    \For{$D_l,D_m \in D$ and $l\neq m$}{
        Compute MB: $C_{MB}$ and $D_u$\;
        \If{$r_{MB}\leq R$ and $A_{ovl}(A,D_u)>0$}{
            Compute: Potential Zone $D_u(l,m)$\;
            $Z \gets Z \cup \{D_u(l,m)\}$\;} 
    }
    \If{$|Z|=0$}{
        \For{$D_l \in D$}{
            Compute: $D_u$
            \If{$A_{ovl}(A,D_u)>0$}{
                Compute: Potential Zone $D_u(l)$\;
                $Z \gets Z \cup \{D_u(l)\}$\;
            }  
        }
    {}%Return $Z_u=Z$} 
    }
    {}%Return $Z_u=Z$}
}
\Else{$M=[M_{ij}]_{|S|\times |Z|}$\;
Update: $M_{ij}=1\;\text{if $s_i$ covered from $z_j\in Z$} $\;
Define: $S(z_j)=\sum_{i=1}^{|S|} M_{ij}$\;
Define: $S(z)=max_{z_j\in Z}(S(z_j))$\;
Return $Z_u=\{z_j \in Z :\;S(z_j)=S(z)\}$\;} 
\end{algorithm}

Let $S'$ be the set of all eMBB users that are being served by $u$ at $t=t_0$ and their corresponding positions are given by
%\vspace{-2mm}
\begin{equation}
    e'_i=(x'_i,y'_i), \; i=1,\cdots,|S'|.
    %\vspace{-3mm}
\end{equation}
%respectively, where $|S|+|S'|\leq l$.
%\vspace{-8mm}
% \begin{algorithm}
%     \caption{Final Position Finding Algorithm}
%     \label{min_handoff_emBB1}
%     \KwIn{$Z_u, S'$}
%     \KwOut{$z_{i^*}$}
%     \For{$z_i$ in $Z_u$}{ 
%     \For{$s'_i$ in $S_'$}{Calculate $D_o(z_i)$\\
%     \For{$t$ in $D_o(z_i)$ }{ Compute throughput $T(t)$\\
%     $T_{iz_i}=\frac{\sum_{t \in D_o(z_i)}T(t)}{|D_o(z_i)|}$}
%     %Compute $T_{s'_i}$
%     }
%     Compute $T(z_i)=\sum_{j=1}^{|S'|}T_{jz_i}$\\
%     \uIf{$T(z_i) >T(z_{i+1})$}{$z_{i^*}=z_{i}$\\Return $z_{i^*}$}
%     %\uElseIf{Calculate $d(z_i,e'_i)$ and $d(z_i,e'_{i+1})$ }{$d(z)=\min\{d(z_i,e'_i),d(z_i,e'_{i+1})\}$\\
%     %return $z$}
    
%     \Else{
%     \uIf{$T(z_i) = T(z_{i+1})$}{Calculate $d(z_i,e_0)$ and $d(z_{i+1},e_0)$\\
%     $i^*=\argmin\{d(z_{i+1},e_0),d(z_{i+1},e_0)\}$\\
%     return $z_{i^*}$}
%     \Else{$z_{i^*}=z_{i+1}$
%     \\Return $z_{i^*}$}
%     }
%     }
% \end{algorithm}
%\vspace{-2mm}

 Like URLLC users, during $(t_0, t_0+\Delta t)$ interval, the disk within which $s'_i \in S'$ can lie, is given by
 \vspace{-2mm}
 \begin{align}
 %\vspace{-4mm}
 D'_i=&\left\{(x,y)\; :\; (x-x'_i)^2+(y-y'_i)^2\leq v'_i.\Delta t\right\}, \nonumber
 \vspace{-2mm}
 \end{align}
 where $v'_i$ is the velocity of $s'_i$.
 Let us consider that $D_o(iz)=A_{\rm ovl}(D'_i,A_z)$ be the overlapping region between $D'_i$ and $A_z$, where $A_z$ is the circle of radius $R$ centered at $z \in Z_u$. We discretize the region $D_o(iz)$ into uniform cells/grids and define $|D_o(iz)|$ as the total number of grid centers in $D_o(iz)$. The  throughput $T_{iz}$ obtained by $s'_i$ from $z$ is defined as
 \vspace{-1.5mm}
\begin{equation}
      T_{iz}=\frac{\sum_{w \in D_o(iz)}T(wz)}{|D_o(iz)|},
      \vspace{-1.5mm}
  \end{equation}
  %\vspace{-2mm}
where $T(wz)$ is the throughput obtained by $s'_i$ located at $w$ from $u$ located at $z$ and calculated by equation \eqref{throughput}. Note that $T(wz)$  depends on whether $w$ is having LoS or NLoS from $z$. Therefore, the aggregate eMBB throughput obtained from $z$ is now defined as
\vspace{-2mm}
 \begin{equation}
      T(z)=\sum_{i=1}^{|S'|}T_{iz}.
      \vspace{-1mm}
  \end{equation}
Here, we are interested in moving $u$ at $z \in Z_u$ for which the value of  $T(z)$ is maximum and let $z^*$ be such a point, i.e., 
\vspace{-2mm}
\begin{equation}
    T(z^*)=\max_{z\in Z_u} T(z).
\end{equation}
Let $Z'_u=\{z \in Z_u: T(z) = T(z^*)\}$. Note that if $|Z'_u| > 1$, we select that $z \in Z'_u$ for which the displacement $d(e_0,z)$ is minimum. The complete process is described in Algorithm \ref{min_handoff_emBB1}.

\begin{algorithm}[h!]
    \caption{Final Position Finding Algorithm}
    \label{min_handoff_emBB1}
    \KwIn{$Z_u$, $S'$}
    \KwOut{$Z'_{u}$}
    \For{$z$ in $Z_u$}{ 
    \For{$s'_i$ in $S'$}{Calculate $D_o(iz)$\;
    \For{each grid center $w$ in $D_o(iz)$ }{ Compute $T(wz)$ using equation \eqref{throughput}\;
    $T_{iz}=\frac{\sum_{w \in D_o(iz)}T(wz)}{|D_o(iz)|}$\;}
    }
    Compute $T(z)=\sum_{i=1}^{|S'|}T_{iz}$\;
    }
$T(z^*)=\max\left\{T(z)\;|\;z\in Z_u\right\}$\;
Return $Z'_u=\{z \in Z_u\;|\;T(z)=T(z^*)\}$\;
\end{algorithm}
\vspace{-2mm}

\subsection{Complexity Analysis of the Algorithm} 
\noindent A UAV $u$ can provide services to at most $l$ UEs at a particular time slot. That is, $|S|+|S'|\leq l$. Let  $r$ be the maximum disk radius among all the users in $S \cup S'$ and $r_u$ be the radius of disk $A$ corresponding to UAV $u$. Discretizing two circles of radii $r$ and $r_u$ into unit cells and then checking LoS status between every pair of cells, requires $O(r_u^2r^2)$ time complexity. 
\begin{itemize}
\item \textbf{Algorithm 1:} It searches for $Z_u$, by looping over all possible triplets, pairs and individual disks. Also, it needs to check LoS status between every pair of discretized cells to find potential zones. So, worst case time and space complexities are $O(l^3.r_u^2.r^2)$ and $O(r_u^2.l)$ respectively.
%The algorithm searches for $Z_u$, by looping over all possible triplets of disks in $S 3$, and check if $Z_u \neq \phi$ in $O(R^2.r^2)$ time. If passed then update the matrix $M$ in $O(n.R^2.r^2)$ time. If failed then over all possible pairs ($O(l^3)$) of disks in $S_2$ and all individual disks ($O(l)$) in $S_1$  and find $Z_u$ accordingly. So, worst case time complexity is $O(l^3.R^2.r^2)$. Also, the space complexity is $O(R^2.l)$ for the matrix $M$.
\item \textbf{Algorithm 2:} It searches for $Z'_u \subset Z_u$ by calculating the throughput for every users in $S'$ and also checking the LoS status between every pair discretized cells. Hence worst case time and space complexities are $O(l.r_u^2.r^2)$ and $O(r_u^2.l)$ respectively.

\end{itemize}

\begin{figure*}[t]
\begin{subfigure}[b]{.33\textwidth}
    \centering
    \includegraphics[width=1\linewidth]{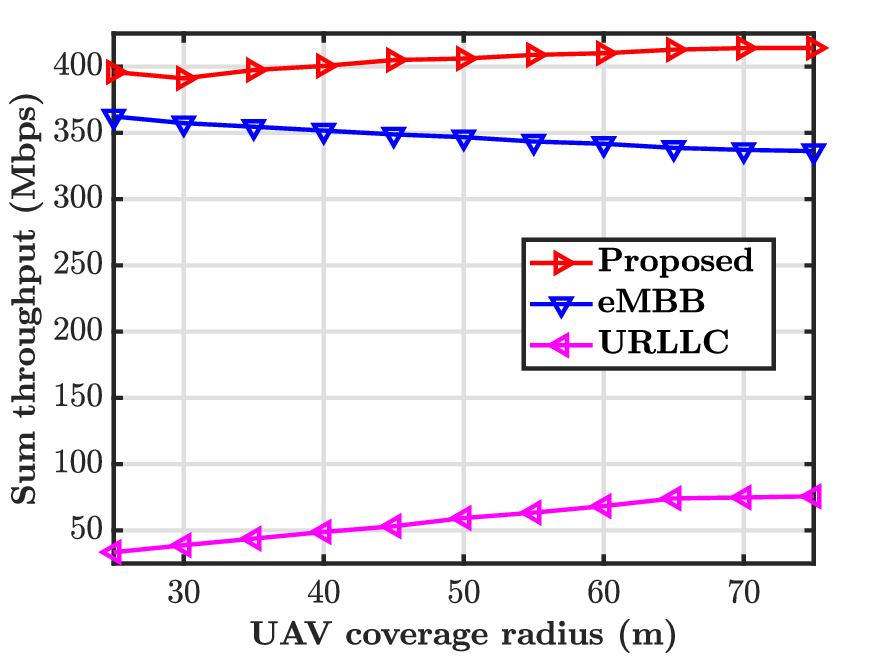}
    \vspace{-6mm}
    \caption{}
    %\caption{Patch spacing vs achievable data rate.}
    \vspace{-2mm}
    \label{patch distance}
\end{subfigure}
 \begin{subfigure}[b]{.33\textwidth}
    \centering
    \includegraphics[width=1\linewidth]{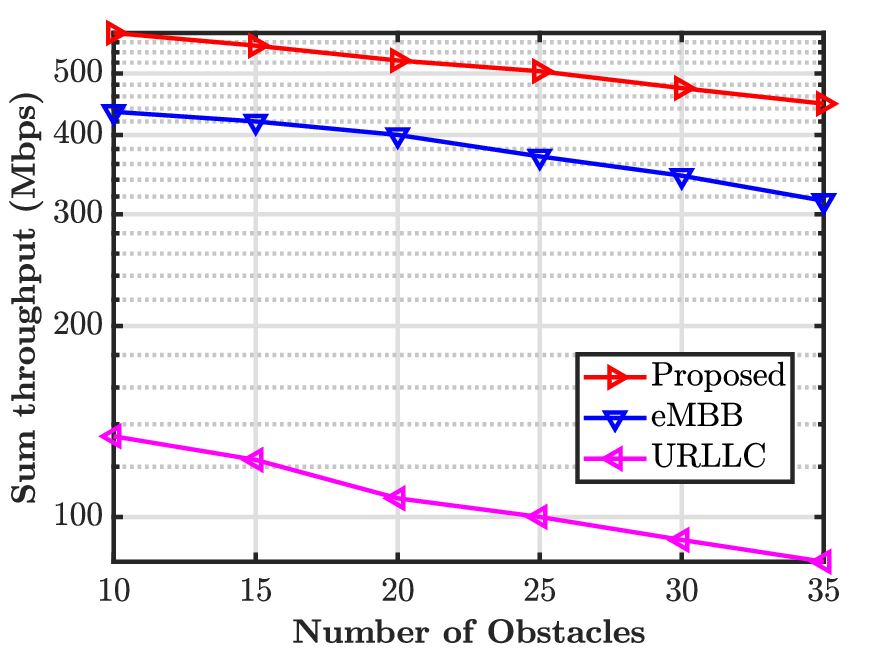}
    \vspace{-6mm}
    \caption{}
    %\caption{No. of groups vs achievable data rate.}
    \vspace{-2mm}
    \label{data rate vs no of groups}
\end{subfigure}
\begin{subfigure}[b]{.33\textwidth}
    \centering
    \includegraphics[width=1\linewidth]{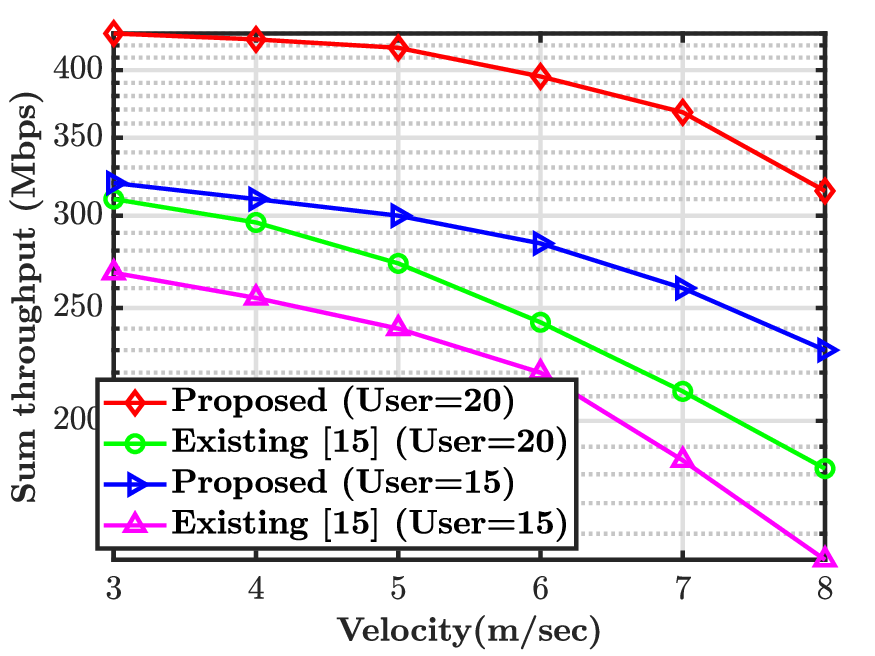}
    \vspace{-6mm}
     \caption{}
    %\caption{No. of patches in a group vs achievable data rate.}
    \vspace{-2mm}
    \label{phase shift}
\end{subfigure}
\caption{\footnotesize  Impact of (a) Coverage radius of UAV, (b) Number of obstacles, and (c) User velocity on Sum throughput.}
\vspace{-6mm}
\end{figure*}
\vspace{-.5em}
\section{Simulation Results}\label{simulation}
\noindent In this section, our main goal is to assess how well the suggested approach works in different environmental settings. We have included a performance comparison study against an existing approach \cite{existing}. In \cite{existing}, a trajectory planning is obtained by  taking into account the mobility of users and the obstacles in their environment, without considering the influence of user traffic characteristics. The objective of our proposed approach is to optimize the sum throughput obtained by URLLC and eMBB users while serving a maximum number of URLLC users. The simulation parameters that we used are given in Table \ref{sim_params_table}

\begin{table}[H]
\centering
\small  % or \footnotesize
\renewcommand{\arraystretch}{1.2} % Slightly tighter row spacing
\begin{tabular}{|c|c|}
\hline
\textbf{Parameter} & \textbf{Value} \\
\hline
Communication region & $400m \times 400m$ \\
Coverage Radius ($R$) & $46 m$ \\
Time slot duration ($\Delta t$) & $3$ sec \\
Transmission power $(P_i)$ & $30$ dBm \cite{drams_lakshmi_da} \\
URLLC threshold ($R_{\rm req}$) & 10 Mbps [ITU \& 3GPP] \\
Rician fading factor ($K$) & $2$ \cite{drams_lakshmi_da} \\
Altitude of UAV & [$22$ m, $150$ m] \cite{geometric} \\
Carrier frequency & $73$ GHz \cite{paramss} \\
LoS link parameters & $\alpha = 69.8$, $\beta = 2$, $\sigma_0=3.1$ \cite{paramss} \\
NLoS link parameters & $\alpha = 82.7$, $\beta = 2.69$, $\sigma_0=8.7$ \cite{paramss} \\
\hline
\end{tabular}
\caption{Simulation Parameters}
\label{sim_params_table}
\end{table}
\vspace{-1em}  

Fig.\ref{patch distance} shows a comparison of eMBB throughput, URLLC throughput, and sum throughput obtained by our proposed algorithm for different UAV coverage radii. Here eMBB and URLLC throughputs represent the aggregate throughput obtained by the eMBB and URLLC users respectively. Sum throughput (Proposed) represents the sum of eMBB and URLLC throughputs. Note that the coverage radius of the UAV is increased in Fig \ref{patch distance} up to the limit such that the minimum data rate requirement of the URLLC users is satisfied. Hence the URLLC throughput increases with the increase of UAV coverage radius. On the contrary, as the UAV coverage radius increases, throughputs obtained by the eMBB users decrease as achievable throughput is inversely proportional to the distance from the user to the UAV. Since URLLC is highly prioritized, the increment of URLLC throughput is slightly more than the decrement of eMBB throughput. Hence sum throughput has an increasing trend.

Furthermore, Fig.\ref{data rate vs no of groups} investigates the impact of obstacle density on the sum throughput, eMBB throughput, and URLLC throughput for a fixed number of users ($35$). As obstacle density increases, the possibility of having LoS decreases. Thus all the throughputs decrease. Note that URLLC users are considered to be covered if all the points in its disk get LoS and eMBB users can be covered partially. That's why we observe that URLLC throughput decreases more rapidly than the eMBB throughput.  Accordingly, the sum throughput is also decreasing as a combined effect of both of them.

Finally, Fig.\ref{phase shift} presents a comparative analysis between the proposed strategy and the existing scheme \cite{existing} under varying user velocities, for a fixed number of users ($15$ and $20$). It is observed that our approach significantly outperforms the method described in \cite{existing}. The key reason behind this improvement is the differentiated treatment of URLLC and eMBB users based on their inherent service requirements, unlike \cite{existing} which treats all users uniformly. Specifically, due to the ultra-reliability and low-latency demands of URLLC, our method first prioritizes finding a zone from where $u$ serves as many URLLC users as possible. Subsequently, from that zone, we aim to find UAV positions that maximize the service for eMBB users by leveraging partial coverage, acknowledging that eMBB users can tolerate moderate latency or brief service interruptions. In contrast, the approach in \cite{existing} treats all users as URLLC, meaning that eMBB users also receive low-latency, ultra-reliable service, resulting in inefficient use of network resources. This performance gap is clearly reflected in the figure. Moreover, we observe that our strategy yields more consistent results than \cite{existing} across varying user velocities, demonstrating its robustness and adaptability to different environmental conditions.

\vspace{-.5em}
\section{Conclusion}\label{conclusion}
\noindent In this work, we investigated the impact of user and UAV mobility on system performance considering the varying user traffic characteristics in the air-to-ground communication framework. Based on traffic characterization and the importance of LoS links, we proposed a UAV trajectory planning algorithm. In this context, we have used the concept of Apollonius circle and minimum enclosing ball of balls to find the optimal UAV location. The numerical results demonstrated the superiority of our proposed framework over an existing benchmark scheme.

\vspace{-.5em}
\bibliographystyle{IEEEtran}
\bibliography{main} 

\end{document}